\documentclass[conference]{IEEEtran}
\IEEEoverridecommandlockouts
\usepackage{cite}
\usepackage{amsmath,amssymb,amsfonts}
\usepackage{algorithmic}
\usepackage{graphicx}
\usepackage{textcomp}
\usepackage{multirow}
\usepackage{dsfont}
\usepackage{bbm}
\usepackage{pifont}
\usepackage{xcolor}
\usepackage[T1]{fontenc}
\usepackage{aecompl}
\usepackage{threeparttable}
\def\BibTeX{{\rm B\kern-.05em{\sc i\kern-.025em b}\kern-.08em
    T\kern-.1667em\lower.7ex\hbox{E}\kern-.125emX}}
    
\begin{document}
% \linenumbers
\title{SF-IDS: An Imbalanced Semi-Supervised Learning Framework for Fine-grained Intrusion Detection\\
\thanks{
\IEEEauthorrefmark{1}Contributed Equally \IEEEauthorrefmark{3}Corresponding Author
}
}

\author{
\IEEEauthorblockN{
Xinran Zheng\IEEEauthorrefmark{2}\IEEEauthorrefmark{1},
Shuo Yang\IEEEauthorrefmark{2}\IEEEauthorrefmark{1}, and
Xingjun Wang\IEEEauthorrefmark{2}\IEEEauthorrefmark{3}}
\IEEEauthorblockA{\IEEEauthorrefmark{2}Tsinghua Shenzhen International Graduate School, Tsinghua University, Shenzhen, China}
}

\maketitle

\begin{abstract}
Deep learning-based fine-grained network intrusion detection systems (NIDS) enable different attacks to be responded to in a fast and targeted manner with the help of large-scale labels. However, the cost of labeling causes insufficient labeled samples. Also, the real fine-grained traffic shows a long-tailed distribution with great class imbalance. These two problems often appear simultaneously, posing serious challenges to fine-grained NIDS. In this work, we propose a novel semi-supervised fine-grained intrusion detection framework, SF-IDS, to achieve attack classification in the label-limited and highly class imbalanced case. We design a self-training backbone model called RI-1DCNN to boost the feature extraction by reconstructing the input samples into a multichannel image format. The uncertainty of the generated pseudo-labels is evaluated and used as a reference for pseudo-label filtering in combination with the prediction probability. To mitigate the effects of fine-grained class imbalance, we propose a hybrid loss function combining supervised contrastive loss and multi-weighted classification loss to obtain more compact intra-class features and clearer inter-class intervals. Experiments show that the proposed SF-IDS achieves 3.01\% and 2.71\% Marco-F1 improvement on two classical datasets with 1\% labeled, respectively.
\end{abstract}


\section{Introduction}
The fine-grained network intrusion detection system (NIDS) can help experts to take targeted measures to address the impact of cyber attacks. Deep learning has brought significant performance improvements to existing systems with continuous development, but it relies heavily on large-scale labeled samples. This leads to two persistent and often co-occurring general challenges. The first one is the lack of labeled samples. Labeling data is always a costly task and often requires the support of domain experts. The second one is real fine-grained traffic often tends to show a long-tailed distribution with severe class imbalance, which creates ``label bias'' during training, making the decision boundary driven by the head class, resulting in poor classification performance.

Several works \cite{relateMSML,relatemixup,relateDQN} related to the above challenges are proposed, but they focus on one of them while assuming the other does not exist, and few expect to address both simultaneously. Semi-supervised learning is a common approach used in the case of insufficient labeled samples. A typical idea is self-training, which generates pseudo-labels for unlabeled samples to expand the labeled dataset. Highly class imbalance reduces the accuracy of pseudo-labels on minority classes, poisoning the labeled samples. For the second challenge, current works have researched the class imbalance problem under supervised learning. These methods rely on label information to rebalance the data, and using them only for limited labeled samples may cause overfitting. Also, the long-tailed shape of fine-grained traffic sample distribution is more challenging than the plain class imbalance (e.g., imbalance in binary classification), which has not been explored in depth, especially in semi-supervised learning scenarios.  

Based on the above issues, we propose SF-IDS to achieve fine-grained intrusion detection in the case of insufficient labeled samples and high-class imbalance. Specifically, SF-IDS uses self-training to flexibly exploit the supervised value of unlabeled samples and their pseudo-labels. The RI-1DCNN is proposed as a backbone model to enhance the feature extraction capability by reconstructing the input samples into multichannel images. In addition, the uncertainty and prediction probability of pseudo-labels is used as the basis for label filtering. To address the extreme class imbalance problem, we design a hybrid loss function with a combination of supervised contrastive learning and multi-weighted classification loss to enable good feature representation to correct biased classifiers. The prototype system of this framework is implemented in this paper, and the major contributions are the following four folds:

\begin{itemize}
% \item{We propose a novel imbalanced semi-supervised learning framework, SF-IDS, for NID. This is the first time that a self-training based semi-supervised learning is used to achieve long-tailed attack classification with the lack of labeled samples.}
\item{We propose SF-IDS to achieve fine-grained network intrusion detection. This is the first time that self-training-based semi-supervised learning is used in NIDS to address both the labeled sample shortage and long-tailed attack classification challenges.}
\item{We design RI-1DCNN as the backbone model of SF-IDS to enhance the traffic feature extraction ability by reconstructing multichannel images in training. An uncertainty-based label filter is proposed to mitigate pseudo-label noise by evaluating uncertainty and resampling.}
\item{We design a hybrid loss combining supervised contrastive loss and multi-weighted classification loss as the learning objective to obtain tighter intra-class features and clearer classification boundaries under severe class imbalance.}
\item{We evaluate the proposed model on two classical datasets: NSL-KDD and CICIDS2017. The experimental results show that SF-IDS still has excellent fine-grained attack classification capability with only 1\% labeled samples.}
\end{itemize}
% The label uncertainty is evaluated as the extra reference for pseudo-label filtering to mitigate the label noise.
% \vspace{-2mm}

\begin{figure*}[ht]
\centering
\includegraphics[width=0.9\linewidth]{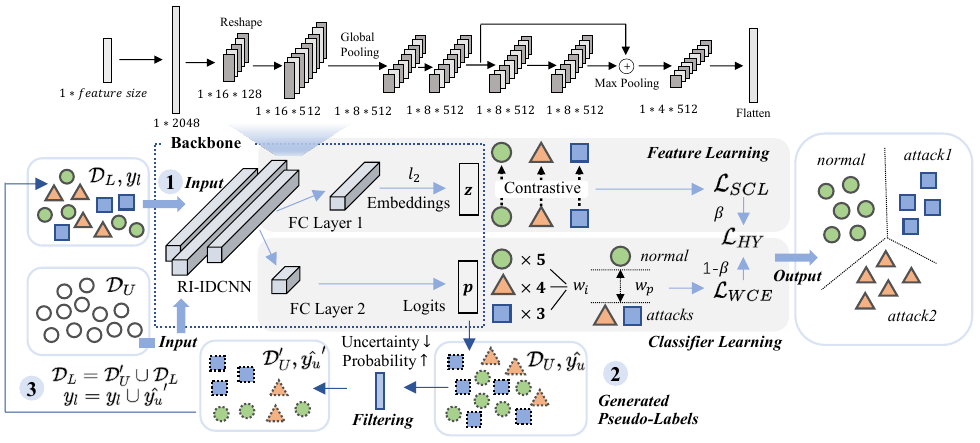}
\caption{The overview of the proposed imbalanced semi-supervised learning framework, SF-IDS.}
\label{Arch}
\end{figure*}

\section{Related Work}

Network intrusion detection system (NIDS) has been receiving a lot of attention over the past decade. Some early works focus on traditional machine learning, such as random forest\cite{relaterf}, SVM\cite{relatesvm}, and decision tree\cite{relatedt}, whose performance depends on effective feature engineering. As attackers continuously update their behavior, finding adaptive features becomes difficult. Recently, the automatic feature extraction ability of deep learning has received increasing attention, KNN\cite{relateknn}, DNN\cite{relatednn}, RNN\cite{relaternn}, and other models have been used in the field of intrusion detection. Imrana \textit{et al.}\cite{relatebilstm} uses BiLSTM to focus on more contextual information and solve the gradient disappearance problem to improve the detection accuracy of two minority attacks, U2R and R2L. Sahu \textit{et al.}\cite{relatelstm} proposes LSTM combined with FCNN for extracting Spatio-temporal features of traffic achieves good results in six datasets on intrusion detection. The good performance of these works is usually based on two assumptions: sufficient labeled samples and a class-balanced dataset. However, this often cannot be satisfied simultaneously in practical applications.

Zhang \textit{et al.}\cite{relateSMOTE} proposes the model aiming to solve the class imbalance problem under supervised learning, which combines SMOTE model and under-sampling for clustering based on Gaussian Mixture Model (GMM) to solve the class imbalance problem by resampling. Zhang \textit{et al.}\cite{relatePCCN} proposes parallel cross convolutional neural network to extract feature representations and avoid the neglect of few-sample categories. These commonly used methods often require labels as the basis for rebalancing the dataset or feature learning, which may lead to overfitting in case of insufficient labels.

Semi-supervised learning is often used to overcome the lack of labeled samples. Among them, self-training is a competitive method that can flexibly adapt to data of different sizes \cite{relaterethinking} and generate pseudo-labels for unlabeled samples thus achieving the performance goals of supervised learning. Hou \textit{et al.} \cite{relatemixup} uses \textit{mixup} to generate augmented samples and filters reliable pseudo-labels based on consistency. Decision trees are used as classification models. Li \textit{et al.} \cite{expersemi} proposes a semi-supervised framework, Semi-WTC, which uses active learning to extract key unlabeled data for manual labeling and artificially assigns modified feature weights to hard-to-score samples. However, the class imbalance problem is still the reason for the poor accuracy of existing methods in minority classes, especially in fine-grained classification tasks. Severe class imbalance leads to biased pseudo-labels, which further degrades the performance of the model.

In fact, there is a relative lack of methods that enable semi-supervised fine-grained intrusion detection while addressing the insufficient labels and class imbalance. Therefore, the goal of our work is precisely to propose an efficient framework that can simultaneously solve the insufficient labels and imbalances in fine-grained attack classification.

% Therefore, the goal of our work is precisely to propose an efficient framework to achieve fine-grained attack classification with limited labeled samples and long-tailed distribution of classes.
% \cite{relateReview}

\section{Proposed Model}

\subsection{Overview}

This section discusses the proposed imbalanced semi-supervised learning framework, SF-IDS, and how it copes with the lack of labeled samples and the imbalanced class distribution of fine-grained attacks.
% This section proposes SF-IDS as a novel imbalanced semi-supervised learning framework for NID, addressing both the lack of labeled samples and the imbalanced class distribution of fine-grained attacks. $\hat{\mathcal{L}}=\left\{\left(\mathcal{D}_L, y_l\right)\right\}$ $\left(\mathcal{D}_U^{\prime}, \hat{y}_{u}^{\prime}\right)$
Fig. \ref{Arch} illustrates the overall architecture of the framework. SF-IDS implements three main steps of self-training using a newly designed backbone model RI-1DCNN. First, RI-1DCNN is trained by original labeled samples. Later, the above model is used to generate pseudo-labels $\hat{y}_{u}$ for unlabeled samples $\mathcal{D}_U$. The uncertainty of the pseudo-labels is evaluated at the same time as one of the references for mitigating the label noise. After filtering, some of the pseudo labels ${\hat{y}_{u}}^{\prime}$ are retained. Finally, the original labeled samples and filtered pseudo-labeled samples are combined as new input to retrain the model. The second and third steps are performed several times to update pseudo-labels, allowing the model to gain stronger generalization. We propose a hybrid loss as the optimization objective. Supervised contrastive loss and multi-weighted cross-entropy loss are jointly optimized so that good features guide unbiased classifiers.

% , which gives the model superior classification ability with fine-grained class imbalance.

\subsection{RI-1DCNN}
% we use RI-1DCNN as its central underlying classifier
The CNN model is suitable for extracting feature associations to generate good representations, which helps to aggregate intra-class features more accurately. Generally, it performs well in image processing, while the processed traffic data is similar to tabular data, which limits its capability. Therefore, we propose the RI-1DCNN as a self-training backbone model to reconstruct the input samples into a multichannel image format. We first add a fully-connected (FC) layer before the convolutional layers to extend the input features to provide enough virtual pixels for subsequent operations. The back-propagation process allows the FC layer to learn the correct feature ordering and thus give a specific meaning to the image. Then, the features are reshaped into multi-channel images, five convolutional layers are used to extract fine feature associations. A residual connection is added to lightweight the model and avoids gradient disappearance. To jointly learn the feature extractor and classifier, RI-1DCNN has a projection head and a classification head. The projection head is a narrower hidden layer, which maps the sample representation $\boldsymbol{r}$ into a lower dimension vector $\boldsymbol{z}=\operatorname{Proj}(\boldsymbol{r}) \in \mathcal{R}^{D_P}$ of size $D_P$ for generating feature embeddings. Then, $\ell_2$ normalization is applied to $\boldsymbol{z}$ so that the inner product can be used as distance measurements. The classification header is used to output the gradients to evaluate the multi-weight classification loss.

\subsection{Uncertainty-based Label Filter}\label{Filter}
Pseudo-label noise affects the performance of self-training. A common approach is to use the prediction probability as the confidence to select reliable labels. However, the prediction probability does not fully reflect the reliability of the classification.\cite{iclruncertion} Thus, we propose an uncertainty-based label filtering strategy. The uncertainty of the pseudo-labels and the prediction probability are combined as filtering references. To be specific, we use MC-Dropout \cite{Uncertainty} to obtain an uncertainty measure by calculating the standard deviation of $T$ stochastic forward passes. The random stops of each neuron of dropout layers conform to the Bernoulli distribution, so the predicted probability can be expressed as the following equation:

\begin{equation}
\label{eq6}
\tilde{p}_{ic} \approx \frac{1}{T} \sum_{t=1}^T softmax\left(f^{\hat{W}_t}(x_i)\right),
\end{equation}
where $\hat{\mathrm{W}}_{\mathrm{t}}$ denotes the model parameters at each sampling, and $f(\cdot)$ represent the model. By definition, the uncertainty of the label is shown as follows:

\begin{equation}
\label{eq7}
u\left(\tilde{p}_{ic}\right) = \sqrt{\frac{1}{T} \sum_{t=1}^{T}\left(f^{\hat{W}_{t}}(x_i)-\tilde{p}_{ic}\right)^2}.
\end{equation}

We assume that reliable pseudo-labels have low uncertainty and high predictive probability. $\tilde{f}_{i}$ is used to denote the filter that the pseudo-label $y_{i}$ of a sample $x_i$ is considered reliable only if $\tilde{f}_{i}$ of that sample is not equal to 0.
\begin{equation}
\label{eq8}
\tilde{f}_{i}=\mathds{1}\left[u\left(\tilde{p}_{ic}\right) \leq \kappa_p\right] * \mathds{1}\left[\tilde{p}_{ic} \geq \tau_p\right],
\end{equation}  
where $\kappa_p$ and $\tau_p$ denote the thresholds of uncertainty and probability, respectively. Not all pseudo-labels are put back into the original training set, as this may lead to an increase in class imbalance and neglect of hard-to-classify samples. We use Borderline-SMOTE\cite{smote} to correct the pseudo-label imbalance problem by generating some samples near the classification boundary. In order not to destroy the learned feature representations in the imbalance case, we resample the pseudo-labels in a form close to the labeled sample distribution and control the degree of class imbalance (the ratio of the number of samples from the largest class to the smallest class) of the pseudo-labeled dataset within a certain threshold.

\subsection{Hybrid Loss}\label{TDLoss}

By rethinking the results of self-training based semi-supervised NIDS, we identify the key factors affecting the performance: 1) The class imbalance causes the feature distribution learned from typical cross-entropy can be highly skewed \cite{TDlossfeature}. This leads to biased classifiers and obtaining the incorrect pseudo-labels. 2) Some hard-to-classify attack samples tend to exhibit similar patterns to normal attack samples, which makes the feature distributions close and the model difficult to obtain clear classification boundaries. Based on the above issues, we propose a hybrid loss function consisting of supervised contrastive loss and multi-weighted cross-entropy loss to obtain compact intra-class features and clear inter-class boundaries to improve attack classification.

\paragraph{Supervised Feature Contrastive Loss} Inspired by\cite{TDlossSupcon}, we construct supervised contrastive loss for intrusion detection tasks to learn better feature representations from imbalanced data. The output of the projection head in RI-1DCNN is the feature $z_{i}$ of the anchor $x_{i}$. All samples with the same label as $x_{i}$ are considered as positive pairs, which are defined as $\left\{\mathbf{x}_i^{+}\right\}=\left\{\mathbf{x}_j \mid y_j=y_i, i \neq j\right\}$. $\left\{\mathbf{z}_i^{+}\right\}$ represents the set of their feature vectors. $y_i$ is the label of sample $x_{i}$. The following equation shows the supervised contrastive loss:

\begin{equation}
\label{eq1}
\mathcal{L}_{S C L}=\sum_{i=1}^{N_{b}} \mathcal{L}_i,
\end{equation}

\begin{equation}
\label{eq2}
\mathcal{L}_i =\frac{-1}{\left|\left\{\mathbf{z}_i^{+}\right\}\right|} \sum_{\mathbf{z}_j \in\left\{\mathbf{z}_i^{+}\right\}} \log \frac{\exp \left(\mathbf{z}_i \cdot \mathbf{z}_j / \tau\right)}{\sum_{\mathbf{z}_k, k \neq i} \exp \left(\mathbf{z}_i \cdot \mathbf{z}_k / \tau\right)},
\end{equation}
where $N_{b}$ denotes the batch size. $\tau \in \mathcal{R}^{+}$ is a scalar temperature parameter used to adjust the model's focus on the distance between samples. $\mathcal{L}_{SCL}$ computes a weighted average of the similarity between $x_i$ and its all positive pairs, flexibly including an arbitrary number of positive samples, and optimizing the agreements between them.

\paragraph{Multi-weighted Classification Loss} 
The intrusion detection dataset has the high-class imbalance and contains low-frequency classes with small sample sizes. Directly using the sample number proportion of classes as classification loss weights may lead to overfitting and corrupting the original feature representations. For this, we designed a smooth class imbalance weight $w_{i}$:

\begin{equation}
\label{eq3}
w_{i}=\frac{\log \left(N_{\min }+n\right)}{\log \left(N_i+n\right)},
\end{equation}
where $N_{\min }$ is the number of samples in the smallest class and $N_i$ is the number of samples in the class $i$. The constant $n$ is used to compensate the low-frequency classes to prevent over-correction, and $log$ is for smoothing the distribution.

The confusion of normal and attack samples also affects the performance of classification. We propose probabilistic reset weights $w_p$ to make the model more focused on such mistakes. All types of attack traffic are considered as a uniform anomaly class $\mathcal{A}$ and the class of normal samples is $\mathcal{N}$. According to the error type of the prediction, the original gradient of the corresponding sample is fine-tuned. Equation \ref{eq4} shows the calculation of the weights.

\begin{equation}
\label{eq4}
w_{pi} = \mathds{1}_{\hat{y_{i}} \neq y_{i}}\mathds{1}_{\hat{y_{i}}\in \mathcal{N}} * \alpha+\mathds{1}_{\hat{y_{i}} \neq y_{i}}\mathds{1}_{\hat{y_{i}}\in \mathcal{A}} * \alpha,
\end{equation}
where $\hat{y_{i}}$ denotes the predicted label and $y_{i}$ denotes the true label. When the model confounds the normal and abnormal sample, the parameter $\alpha$ is activated to adjust the model's attention to such misclassification. The weighted classification loss $\mathcal{L}_{WCE}$ is calculated as follows:

\begin{equation}
\label{eq5}
\mathcal{L}_{WCE}=-\frac{1}{K}\sum_{i=1}^K \sum_{c=1}^M w_i g_{ic} \log \left(w_{pi}p_{i c}\right),
\end{equation}
where $K$ and $M$ denote the number of samples and categories, respectively, and $g_{ic}=\{0,1\}$ is the sign function, which takes the value 1 if sample $i$ is in the correct category $c$. $p_{ic}$ is the probability that sample $i$ is in category $c$. 

\paragraph{Calculate hybrid loss}
The hybrid loss $\mathcal{L}_{HY}$ combines supervised contrastive loss $\mathcal{L}_{SCL}$ and multiple weighted classification loss $\mathcal{L}_{WCE}$:

\begin{equation}
\label{eq9}
\mathcal{L}_{HY}=(1-\beta)*\mathcal{L}_{WCE}+\beta * \mathcal{L}_{SCL}.
\end{equation}

The model should focus more on obtaining a good feature representation in the early stages of training, and on improving the classification performance in the later stage. Thus, we use the variable parameter $\beta$, which varies inversely with epoch, to adjust the weights of different losses, making better feature learning to help simplify the training of the classifier.

% Better feature learning can help simplify the training of classifiers. Thus $\beta$ is a weighting factor that is inversely proportional to the epoch and is used to adjust the relative weights of the classification and feature learning.

% Thus, the multi-weight classification loss $\mathcal{L}_{WCE}$ for unlabeled samples can be represented as:
% \begin{equation}
% \label{eq8}
% \mathcal{L}_{WCE}=-\frac{1}{N} \sum_{i=1}^K \sum_{c=1}^M w_i \tilde{y}_{ic} \log \left(w_p \tilde{p}_{ic}\right)
% \end{equation} 

\section{Experimental Result}

In this section, we design three experiments to evaluate the performance of the SF-IDS. First, we validate the fine-grained classification performance with 1\% labeled samples in different datasets. Besides some classical supervised models, we choose the well-performing semi-supervised model FixMatch\cite{experfixmatch} and the state-of-the-art semi-supervised intrusion detection model Semi-WTC\cite{expersemi} as the baselines. The results indicate that SF-IDS improves all metrics, especially \textit{Marco-F1}, and has generalizability across different datasets. Second, we evaluated the performance with different label ratios and different unlabeled sample sizes, and SF-IDS obtained a boost in \textit{precision} and \textit{Marco-F1}. Finally, a simple ablation experiment validates the effectiveness of the proposed hybrid loss. The experiments are implemented using the deep learning framework Pytorch and run on a server with GeForce RTX 3090Ti and Intel(R) Xeon(R) Gold 6230R 2.10GHz CPUs.

\begin{table}
\caption{Dataset Description}
\label{dataset}
\centering
\begin{threeparttable} 
\begin{tabular}{|c|c|c|c|c|c|c|} 
\hline
\multirow{2}*{\textbf{No.}} & \multicolumn{3}{c|}{\textbf{NSL-KDD}} & \multicolumn{3}{c|}{\textbf{CICIDS2017 }} \\ 
\cline{2-7} & \textbf{Class} &\textbf{Train}& \textbf{Test} & \textbf{Class} & \textbf{Train} & \textbf{Test}  \\ 
\hline
0 & normal& 616  & 15411  & BENIGN  & 16284  & 407101\\ 
\hline
1  & neptune & 367 & 9174& Hulk & 1372  & 34302 \\ 
\hline
2  & attack  & 51 & 1267 & DDoS & 1024 & 25601 \\ 
\hline
3  & satan  & 35  & 874 & PortScan & 458  & 11461 \\ 
\hline
4   & ipsweep & 30 & 748 & GoldenEye  & 82  & 2059  \\ 
\hline
5   & smurf & 26   & 662& FTP-Patator & 44  & 1096 \\ 
\hline
6   & portsweep & 25  & 618  & slowloris & 42  & 1058   \\ 
\hline
7  & nmap & 13  & 313   & Slowhttptest & 41  & 1035 \\ 
\hline
8  & back & 10 & 263 & SSH-Patator & 25 & 614 \\
\hline
9 & warezmaster & 8  & 193 & Web Attack  & 17 & 419  \\ 
\hline
10 & teardrop & 7  & 181  & Bot & 16  & 389    \\
\hline
\end{tabular}
    \begin{tablenotes}
        \footnotesize
        \item Train denotes the number of labeled training sets.
      \end{tablenotes}
  \end{threeparttable}
\end{table}

\begin{table*}[]
\caption{Result of Comparative Experiments on 1\% labeled NSL-KDD Dataset}
\label{nsl-kdd}
\centering
\begin{tabular}{|c|c|c|c|c|c|c|c|c|c|c|c|c|c|c|c|} 
\hline
 \textbf{Model} & \textbf{Acc.} & \textbf{Pre.} & \textbf{Rec.} & \textbf{F1} & \textbf{0} & \textbf{1} & \textbf{2} & \textbf{3} & \textbf{4} & \textbf{5} & \textbf{6} & \textbf{7} & \textbf{8} & \textbf{9} & \textbf{10}  \\ 
\hline
RF  & 93.58 & 82.27 &82.87 & 80.99 & 91.28 & 97.19 & 57.30 & 80.25 & 67.41& 98.76 & 90.34 &64.92 & 92.07 & 74.12 & 91.34 \\ 
\hline
SVM  &93.41 &83.87 & 81.25 & 82.19 & 90.67 & 99.14 & 56.01    &91.87 & 68.12 & 96.93 & 90.29 & 68.03 &93.79 &  75.11 &92.57 \\ 
\hline
LR  & 89.95 & 84.66 &81.68 & 81.84 & 91.34 & 99.01 & 58.13 
  &92.88 &64.36 & 98.42 & 90.98  & 64.54 & 92.01 &  87.53 & 92.10 \\ 
\hline
VGG16  &95.34 &88.09 &85.14 &86.20 & 96.12 &99.62 &71.67 &82.20  &95.17  & 93.62 &\textbf{98.92}&77.27 & 97.83& 61.82& 94.74\\ 
\hline
DNN  & 94.28 &86.24 &77.94 &80.64 & 94.55 &99.13 & 63.68 & 92.77 & 90.85& 95.52 & 93.55 & 63.53 & 89.11 &73.68 & 92.31   \\ 
\hline
LSTM & 93.94 &87.15 & 83.11 &84.52 & 94.80 &  98.42 &79.85 & 85.98 &91.25 & 94.96 & 85.34 & 62.50 & 93.33 & 78.26& 94.00 \\ 
\hline
FixMatch\cite{experfixmatch}& 95.41 & 91.14 & 90.94 & 90.36 & 91.45 & 96.67 & 76.12 & 90.46 & 90.88 & 98.04 & 95.91  & 80.11 &95.12 & 91.33  & 96.50 \\ 
\hline
Semi-WTC\cite{expersemi} &95.63 & 93.30 & 90.30 & 91.84 & 93.30 &98.67 & 77.99 & \textbf{94.02}  &\textbf{97.44} & 99.06 & 94.67 & 87.73 & 95.16 & \textbf{92.04}  & 96.19  \\
\hline
SF-IDS  &\textbf{97.78} &\textbf{95.95} & \textbf{94.17} &\textbf{94.60} &\textbf{97.48} &  \textbf{99.95} & \textbf{90.54} & 93.78 & 96.44 & \textbf{99.98} &  96.93 & \textbf{92.31} & \textbf{99.89} & 90.84  & \textbf{97.30}  \\
\hline
\end{tabular}
\end{table*}

\begin{table*}[]
\caption{Result of Comparative Experiments on 1\% labeled CICIDS2017 Dataset}
\label{cicids}
\centering
\begin{tabular}{|c|c|c|c|c|c|c|c|c|c|c|c|c|c|c|c|} 
\hline
 \textbf{Model} & \textbf{Acc.} & \textbf{Pre.} & \textbf{Rec.} & \textbf{F1} & \textbf{0} & \textbf{1} & \textbf{2} & \textbf{3} & \textbf{4} & \textbf{5} & \textbf{6} & \textbf{7} & \textbf{8} & \textbf{9} & \textbf{10}  \\ 
\hline
RF & 96.42 &74.70 & 70.88 & 72.56 & 97.78 &90.01 & 84.34 & 78.74  & 79.95 & 91.09 & 85.56 & 80.23 & 92.81 & 15.00 &  26.20 \\ 
\hline
SVM & 96.15 &81.63 &71.98 & 75.27 & 95.07 & 90.16 & 98.89 & 83.01 &  96.70 & 90.58 & 85.85 & 84.66 & 92.80  & 24.01 & 56.14 \\ 
\hline
LR   &96.01 & 77.83 &70.21 &73.01 & 96.48 & 87.97 &96.18 & 85.24 &  91.39 & 89.10  &82.34 & 81.95 & 90.23 &  20.12 &  35.15 \\ 
\hline
VGG16 & 98.47 & 86.36 & 82.68 & 83.61 & 98.66 & \textbf{99.69} & 99.13 & 87.96 & 90.48 & 93.07 & 99.89 & 87.12 &99.98 & 15.38 & 78.57\\ 
\hline
DNN & 97.48 & 85.48 & 75.42 & 77.86 & 98.56 & 99.63 & 94.56 & 83.49& 86.96 & 84.62 &98.69 &90.91& 99.47 & 16.00 & 82.40 \\ 
\hline
LSTM     &98.04 & 85.03 &76.32 &78.91 & 98.49 &97.31 & 97.28 & 82.50 & 86.36 &\textbf{95.83}& 99.98 &83.33 & 85.71 & 14.29 & \textbf{99.89} \\ 
\hline
FixMatch\cite{experfixmatch} &98.23 & 91.60 & 93.90 & 90.26 & 99.01 &96.96 & 99.22 & 85.74   & 90.48 & 90.17 & 99.44& 87.18 &90.30 &   79.01& 90.12 \\ 
\hline
Semi-WTC\cite{expersemi} &98.06 & 92.91 & 93.37 & 91.46 & 98.66 &97.24 & 99.21 & 83.71   & 91.56 & 92.31 & 91.67 & 90.91 &  99.98 & 82.33& 94.41 \\ 
\hline
SF-IDS  &\textbf{98.86} &\textbf{95.77} & \textbf{94.01} & \textbf{93.94} & \textbf{99.23}  & 99.09& \textbf{99.60} & \textbf{90.43}&  \textbf{93.90} & 94.13 & \textbf{99.98} & \textbf{93.49} & \textbf{99.98} & \textbf{87.24} & 96.42 \\
\hline
\end{tabular}
\end{table*}

\subsection{Dataset and Evaluation Measures}

The classic NSL-KDD dataset and the CICIDS2017 dataset were used to comprehensively evaluate the SF-IDS.
\begin{itemize}
    \item{NSL-KDD: This dataset contains 41 features for each sample. It has 77,054 normal and 71,463 abnormal traffic. 80\% of the dataset was selected for training and the remaining 20\% as the test set. 1\% of the training samples are labeled. The categories with too few samples after segmentation are combined together as an additional category called ``attack''. After that, 11 categories are involved in the experiments.}
    \item{CICIDS2017: This dataset is a representation of real network traffic data, which contains 2,830,743 traffic samples, each with 78 attributes. It is extremely unbalanced, with only 24.5\% of the attack samples. Equally, 80\% of the data is used as the training set, with 1\% samples of it labeled. We keep the original fine-grained attack categories as much as possible and merge the categories whose samples are too sparse after segmentation, with 11 categories finally being kept.}
\end{itemize}

We encode the features that are not suitable for the deep learning model and standardize the numerical features. Table \ref{dataset} shows the number of 1\% labeled samples used for training and the size of the test dataset. The evaluation metrics include \textit{Accuracy}, \textit{precision}, \textit{recall}, and \textit{F1-score}, where \textit{precision} and \textit{F1-score} are given more attention in the imbalanced classification task. Notably, we use \textit{Marco-F1} instead of the original \textit{Micro-F1}, which focuses on each category equally and reflect the classification performance more objectively.

\begin{table}[!t]
\caption{semi-supervised classification performance at different label ratios}
\label{different ratio}
\centering
\begin{center}
\begin{tabular}{|c|c|c|c|c|c|} 
\hline
\multirow{3}{*}{\textbf{Dataset}}    & \multirow{3}{*}{\begin{tabular}[c]{@{}c@{}}\textbf{Label} \\\textbf{Ratio}\end{tabular}} & \multicolumn{4}{c|}{\textbf{Precision} / \textbf{Macro-F1}}   \\ 
\cline{3-6} &   &\begin{tabular}[c]{@{}c@{}}\textbf{\textit{L+}}\\\textbf{\textit{0\%UL}}~\end{tabular}  & \begin{tabular}[c]{@{}c@{}}\textbf{\textit{L+}}\\\textbf{\textit{50\%UL}}~\end{tabular}  & \begin{tabular}[c]{@{}c@{}}\textbf{\textit{L+}}\\\textbf{\textit{100\%UL}}~\end{tabular} & \textbf{\textit{Semi-WTC}}   \\ 
\hline
\multirow{3}{*}{CICIDS2017} & 1\%  & \begin{tabular}[c]{@{}c@{}}92.39\\90.91\end{tabular} & \begin{tabular}[c]{@{}c@{}}94.69\\92.08\end{tabular} & \begin{tabular}[c]{@{}c@{}}95.77\\93.94\end{tabular} & \begin{tabular}[c]{@{}c@{}}92.91\\91.46\end{tabular}  \\ 
\cline{2-6}
& 5\%  & 
\begin{tabular}[c]{@{}c@{}}94.41\\92.37\end{tabular}  & \begin{tabular}[c]{@{}c@{}}95.61\\93.61\end{tabular} & \begin{tabular}[c]{@{}c@{}}96.45\\95.01\end{tabular} & \begin{tabular}[c]{@{}c@{}}93.90\\92.39\end{tabular}  \\ 
\cline{2-6}
& 10\%   &
\begin{tabular}[c]{@{}c@{}}95.46\\93.21\end{tabular}  & \begin{tabular}[c]{@{}c@{}}96.62\\94.44\end{tabular} & \begin{tabular}[c]{@{}c@{}}97.48\\95.84\end{tabular} & \begin{tabular}[c]{@{}c@{}}95.34\\93.51\end{tabular}  \\ 
\hline
\multirow{3}{*}{NSL-KDD}    & 1\%  & \begin{tabular}[c]{@{}c@{}}92.83\\92.01\end{tabular}  & \begin{tabular}[c]{@{}c@{}}94.81\\93.97\end{tabular} & \begin{tabular}[c]{@{}c@{}}95.95\\94.60\end{tabular} & \begin{tabular}[c]{@{}c@{}}93.30\\92.01\end{tabular}  \\ 
\cline{2-6}
& 5\%   & 
\begin{tabular}[c]{@{}c@{}}94.99\\94.41\end{tabular}  & \begin{tabular}[c]{@{}c@{}}95.67\\96.16\end{tabular} & \begin{tabular}[c]{@{}c@{}}96.91\\96.75\end{tabular} & \begin{tabular}[c]{@{}c@{}}94.11\\93.13\end{tabular}  \\ 
\cline{2-6} & 10\%  & \begin{tabular}[c]{@{}c@{}}96.47\\96.02\end{tabular}  & \begin{tabular}[c]{@{}c@{}}97.65\\97.18\end{tabular} & \begin{tabular}[c]{@{}c@{}}98.36\\97.93\end{tabular} & \begin{tabular}[c]{@{}c@{}}94.63\\94.21\end{tabular}  \\
\hline
\end{tabular}
\end{center}
\end{table}

\begin{table}[!t]
\caption{Ablation experiments about hybrid loss}
\label{Ablation}
\centering
\begin{center}
\begin{tabular}{|c|c|c|c|c|} 
\hline
$L_{SCL}$  & \ding{53} & $\checkmark$ & \ding{53}  & $\checkmark$  \\ 
\hline
$L_{WCE}$  & \ding{53} & \ding{53}  & $\checkmark$ & $\checkmark$  \\ 
\hline
Pre      & 91.39  & 92.31 & 92.47 & 92.83 \\ 
\hline
Marco-F1 & 90.16  & 91.36 & 91.43 & 92.01 \\
\hline
\end{tabular}
\end{center}
\end{table}

\subsection{Effectiveness of Fine-grained Classification}
The proposed hybrid loss involves the choice of several hyperparameters, among which $\tau$ and $\alpha$ need to be given extra attention due to their impact on fine category classification. We trained the full amount of data several times to determine their values. $\tau$ is used for supervised contrastive loss, the smaller the parameter the more the model focuses on differentiating the most similar difficult samples. But too small the $\tau$ can destroy the learned semantics. Here, $\tau$ is set to $0.05$. $\alpha$ is used to penalize the model for misclassification of normal and attack samples, and the smaller the $\alpha$ the larger the penalty. Thus, we set $\alpha=0.95$ for training. The choice of other hyperparameters is determined according to the dataset and the different demands on the task.

We evaluate the performance of SF-IDS on the NSL-KDD and CICIDS2017 datasets with 1\% labeled samples and present the results in Table \ref{nsl-kdd} and Table \ref{cicids}, respectively. The name of each category name is represented in the column \textit{No.} of the Table \ref{dataset}. \textit{Precision} is used to denote the classification effectiveness for each category. To obtain more objective results, each value is the average of multiple evaluations, expressed as a percentage, and here five are chosen randomly. According to Table \ref{nsl-kdd}, it can be seen that the proposed SF-IDS achieves the best results for the four performance evaluation metrics with only 1\% labeled NSL-KDD dataset. Compared with the optimal comparison models, \textit{precision} improves by 2.84\% and \textit{Marco-F1} improves by 3.00\%. This is due to the fact that SF-IDS makes full use of the value of unlabeled data and combines the hybrid loss function to obtain more compact class features and clearer classification boundaries. Moreover, SF-IDS achieves the best accuracy rate in 7 of the 11 fine-grained attack categories of NSL-KDD. In contrast, some traditional machine learning and supervised models are limited by the labeled sample size, making feature learning difficult. FixMatch requires data augmentation, which may not work well for traffic data. Semi-WTC resamples the data before training, which makes it practically difficult to adapt the model to the extremely unbalanced class distribution. Table \ref{cicids} validates the performance of SF-IDS for fine-grained classification with the lack of labeled samples on the CICIDS2017 dataset. The SF-IDS achieves the best overall metrics with a 3.08\% improvement in \textit{precision} and a 2.71\% improvement in \textit{Marco-F1}. It also has the most SOTA results in the fine-grained classification. The performance on both datasets demonstrates that SF-IDS has good generalization capability.

\subsection{Performance in Different Label Ratios}

% We compare the \textit{precision} and \textit{Marco-F1} of SF-IDS with supervised learning only and add different amounts of unlabeled data at labeling ratios of 1\%, 5\%, and 10\% to validate the fine-grained classification and manual labeling saving ability. 

To further validate the ability of SF-IDS to utilize labeled and unlabeled samples, we compared \textit{precision} and \textit{Marco-F1} for different label ratios (1\%, 5\%, and 10\%) and different unlabeled sample sizes (0\%, 50\%, and 100\%). The best-performing baseline, Semi-WTC, is used as the comparison model. Table \ref{different ratio} shows the results, L and UL denote labeled and unlabeled samples. Even using only labeled samples, the hybrid loss and RI-1DCNN still give SF-IDS excellent classification ability, which makes the pseudo-labels more reliable. In addition, the improvement of the label ratio and the increase of unlabeled samples can improve the metrics. Among them, \textit{Marco-F1} boosts the highest by 3.89\%. Semi-WTC requires artificially modifying the feature weights of hard-to-score samples, lacks the ability to generalize to different data distributions and relies on experience. Meanwhile, its backbone model consists of only some linear layers stacked feature extraction capability is weak.

% To evaluate the ability of SF-IDS to utilize unlabeled data at different annotation ratios, we compare the performance of SF-IDS with supervised learning only and adding different amount of unlabeled data at label ratio of 1\%, 5\% and 10\%. Meanwhile, FixMatch, the best performing of the baseline models, is used to verify the superiority of SF-IDS with different label ratio. The experimental results are presented in Table. Compared with only using supervised learning, SF-IDS achieves a more substantial improvement with limited labels. FixMatch maintains a steady performance improvement, but cannot outperform SF-IDS with a full set of unlabeled samples. 

\subsection{Effectiveness of Hybrid Loss}

A simple ablation experiment is conducted to verify the effectiveness of the hybrid loss $\mathcal{L}_{HY}$ as well as the addition of supervised contrastive loss $\mathcal{L}_{SCL}$ and multi-weighted classification loss $\mathcal{L}_{WCE}$ alone. The backbone model is trained supervised on 1\% labeled NSL-KDD dataset, where $\mathcal{L}_{SCL}$ and $\mathcal{L}_{WCE}$ are alternately turned off and the baseline is the same model with original cross-entropy loss. Table\ref{Ablation} shows the results. $\mathcal{L}_{SCL}$ and $\mathcal{L}_{WCE}$ improve the \textit{Marco-F1} by 1.33\% and 1.41\%, respectively. Experiments show that the hybrid loss tightens the feature distribution of each category by labels, and constructs a clear classification boundary with the help of effective class weights, which improves the classification performance of \textit{Marco-F1} by 2.05\%.

\section{Conclusion}

In this paper, we propose an imbalanced semi-supervised learning framework, SF-IDS, for fine-grained intrusion detection, including a self-training backbone model RI-1DCNN and a hybrid loss, addressing both the insufficient labels and the long-tailed classification problem. The RI-1DCNN reconstructs the input samples as multichannel images by extending the neurons and assigning real meanings to them to enable the convolutional model to better extract feature associations. Uncertainty is used as a reference for pseudo-label filtering to solve the label noise problem. Also, the hybrid loss integrates supervised contrastive learning and multi-weighted classification loss to achieve intra-class feature compactness and classifier correction. SF-IDS has significant performance gains in two datasets with 1\% labeled samples. The framework has the potential to be applied to more cybersecurity domains to solve similar problems in a generalized manner.

\bibliography{cite}

\begin{thebibliography}{10}
\providecommand{\url}[1]{#1}
\csname url@samestyle\endcsname
\providecommand{\newblock}{\relax}
\providecommand{\bibinfo}[2]{#2}
\providecommand{\BIBentrySTDinterwordspacing}{\spaceskip=0pt\relax}
\providecommand{\BIBentryALTinterwordstretchfactor}{4}
\providecommand{\BIBentryALTinterwordspacing}{\spaceskip=\fontdimen2\font plus
\BIBentryALTinterwordstretchfactor\fontdimen3\font minus
  \fontdimen4\font\relax}
\providecommand{\BIBforeignlanguage}[2]{{%
\expandafter\ifx\csname l@#1\endcsname\relax
\typeout{** WARNING: IEEEtran.bst: No hyphenation pattern has been}%
\typeout{** loaded for the language `#1'. Using the pattern for}%
\typeout{** the default language instead.}%
\else
\language=\csname l@#1\endcsname
\fi
#2}}
\providecommand{\BIBdecl}{\relax}
\BIBdecl

\bibitem{relateMSML}
H.~Yao, D.~Fu, P.~Zhang, M.~Li, and Y.~Liu, ``Msml: A novel multilevel
  semi-supervised machine learning framework for intrusion detection system,''
  \emph{IEEE Internet of Things Journal}, vol.~6, no.~2, pp. 1949--1959, 2018.

\bibitem{relatemixup}
Y.~Hou, S.~G. Teo, Z.~Chen, M.~Wu, C.-K. Kwoh, and T.~Truong-Huu, ``Handling
  labeled data insufficiency: Semi-supervised learning with self-training mixup
  decision tree for classification of network attacking traffic,'' \emph{IEEE
  Transactions on Dependable and Secure Computing}, 2022.

\bibitem{relateDQN}
S.~Dong, Y.~Xia, and T.~Peng, ``Network abnormal traffic detection model based
  on semi-supervised deep reinforcement learning,'' \emph{IEEE Transactions on
  Network and Service Management}, vol.~18, no.~4, pp. 4197--4212, 2021.

\bibitem{relaterf}
P.~A.~A. Resende and A.~C. Drummond, ``A survey of random forest based methods
  for intrusion detection systems,'' \emph{ACM Computing Surveys (CSUR)},
  vol.~51, no.~3, pp. 1--36, 2018.

\bibitem{relatesvm}
H.~Wang, J.~Gu, and S.~Wang, ``An effective intrusion detection framework based
  on svm with feature augmentation,'' \emph{Knowledge-Based Systems}, vol. 136,
  pp. 130--139, 2017.

\bibitem{relatedt}
X.-B. Li, ``A scalable decision tree system and its application in pattern
  recognition and intrusion detection,'' \emph{Decision Support Systems},
  vol.~41, no.~1, pp. 112--130, 2005.

\bibitem{relateknn}
Y.~Li, B.~Fang, L.~Guo, and Y.~Chen, ``Network anomaly detection based on
  tcm-knn algorithm,'' in \emph{Proceedings of the 2nd ACM symposium on
  Information, computer and communications security}, 2007, pp. 13--19.

\bibitem{relatednn}
S.~S. Roy, A.~Mallik, R.~Gulati, M.~S. Obaidat, and P.~V. Krishna, ``A deep
  learning based artificial neural network approach for intrusion detection,''
  in \emph{International Conference on Mathematics and Computing}.\hskip 1em
  plus 0.5em minus 0.4em\relax Springer, 2017, pp. 44--53.

\bibitem{relaternn}
M.~Abdel-Basset, V.~Chang, H.~Hawash, R.~K. Chakrabortty, and M.~Ryan,
  ``Deep-ifs: intrusion detection approach for industrial internet of things
  traffic in fog environment,'' \emph{IEEE Transactions on Industrial
  Informatics}, vol.~17, no.~11, pp. 7704--7715, 2020.

\bibitem{relatebilstm}
Y.~Imrana, Y.~Xiang, L.~Ali, and Z.~Abdul-Rauf, ``A bidirectional lstm deep
  learning approach for intrusion detection,'' \emph{Expert Systems with
  Applications}, vol. 185, p. 115524, 2021.

\bibitem{relatelstm}
S.~K. Sahu, D.~P. Mohapatra, J.~K. Rout, K.~S. Sahoo, Q.-V. Pham, and N.-N.
  Dao, ``A lstm-fcnn based multi-class intrusion detection using scalable
  framework,'' \emph{Computers and Electrical Engineering}, vol.~99, p. 107720,
  2022.

\bibitem{relateSMOTE}
H.~Zhang, L.~Huang, C.~Q. Wu, and Z.~Li, ``An effective convolutional neural
  network based on smote and gaussian mixture model for intrusion detection in
  imbalanced dataset,'' \emph{Computer Networks}, vol. 177, p. 107315, 2020.

\bibitem{relatePCCN}
Y.~Zhang, X.~Chen, D.~Guo, M.~Song, Y.~Teng, and X.~Wang, ``Pccn: parallel
  cross convolutional neural network for abnormal network traffic flows
  detection in multi-class imbalanced network traffic flows,'' \emph{IEEE
  Access}, vol.~7, pp. 119\,904--119\,916, 2019.

\bibitem{relaterethinking}
B.~Zoph, G.~Ghiasi, T.-Y. Lin, Y.~Cui, H.~Liu, E.~D. Cubuk, and Q.~Le,
  ``Rethinking pre-training and self-training,'' \emph{Advances in neural
  information processing systems}, vol.~33, pp. 3833--3845, 2020.

\bibitem{expersemi}
Z.~Li, W.~Chen, Z.~Wei, X.~Luo, and B.~Su, ``Semi-wtc: A practical
  semi-supervised framework for attack categorization through weight-task
  consistency,'' \emph{arXiv preprint arXiv:2205.09669}, 2022.

\bibitem{iclruncertion}
\BIBentryALTinterwordspacing
M.~N. Rizve, K.~Duarte, Y.~S. Rawat, and M.~Shah, ``In defense of
  pseudo-labeling: An uncertainty-aware pseudo-label selection framework for
  semi-supervised learning,'' in \emph{International Conference on Learning
  Representations}, 2021. [Online]. Available:
  \url{https://openreview.net/forum?id=-ODN6SbiUU}
\BIBentrySTDinterwordspacing

\bibitem{Uncertainty}
Y.~Gal and Z.~Ghahramani, ``Dropout as a bayesian approximation: Representing
  model uncertainty in deep learning,'' in \emph{Proceedings of the 33rd
  International Conference on International Conference on Machine Learning -
  Volume 48}, ser. ICML'16.\hskip 1em plus 0.5em minus 0.4em\relax JMLR.org,
  2016, p. 1050–1059.

\bibitem{smote}
H.~Han, W.-Y. Wang, and B.-H. Mao, ``Borderline-smote: a new over-sampling
  method in imbalanced data sets learning,'' in \emph{International conference
  on intelligent computing}.\hskip 1em plus 0.5em minus 0.4em\relax Springer,
  2005, pp. 878--887.

\bibitem{TDlossfeature}
C.~Huang, Y.~Li, C.~C. Loy, and X.~Tang, ``Learning deep representation for
  imbalanced classification,'' in \emph{2016 IEEE Conference on Computer Vision
  and Pattern Recognition (CVPR)}, 2016, pp. 5375--5384.

\bibitem{TDlossSupcon}
P.~Khosla, P.~Teterwak, C.~Wang, A.~Sarna, Y.~Tian, P.~Isola, A.~Maschinot,
  C.~Liu, and D.~Krishnan, ``Supervised contrastive learning,'' \emph{Advances
  in Neural Information Processing Systems}, vol.~33, pp. 18\,661--18\,673,
  2020.

\bibitem{experfixmatch}
K.~Sohn, D.~Berthelot, N.~Carlini, Z.~Zhang, H.~Zhang, C.~A. Raffel, E.~D.
  Cubuk, A.~Kurakin, and C.-L. Li, ``Fixmatch: Simplifying semi-supervised
  learning with consistency and confidence,'' \emph{Advances in neural
  information processing systems}, vol.~33, pp. 596--608, 2020.

\end{thebibliography}

\vspace{12pt}

\end{document}